# Comment on "Nonreciprocal Light Propagation in a Silicon Photonic Circuit" in *Science* 333, 729-733 (2011)

**Alexander Petrov***, **Dirk Jalas and Manfred Eich**

*Hamburg University of Technology, Institute of Optical and Electronic Materials, Eissendorfer Str. 38, D-21073 Hamburg, Germany*

**Michael Krause, Hagen Renner, Ernst Brinkmeyer**

*Hamburg University of Technology, Institute of Optical Communication Technology, Eissendorfer Str. 40, D-21073 Hamburg, Germany*

**Christopher R. Doerr**

*Bell Laboratories, Alcatel-Lucent, Holmdel, NJ 07733 USA*

*Corresponding author: a.petrov@tuhh.de

**Abstract:** In the recent article by Feng et al. [Science **333**, 729-733 (2011)] a system is proposed with a periodic spatial modulation of refractive index and absorption. It is claimed that this system is nonreciprocal and can be used as an optical isolator. The authors use an unconventional definition of reciprocity and thus make wrong conclusions on the applicability of the described phenomena for optical isolators. (Submitted to *Science* on October 7, 2011)

It is well known that the Lorentz reciprocity theorem applies to systems made of inhomogeneous linear lossless or lossy materials, provided only that all permeability and permittivity tensors are symmetric and do not change with time.[1] The system presented in the article by Feng et al. [Science **333**, 729-733 (2011)] fulfills all of these preconditions. Thus, it cannot per se demonstrate nonreciprocal behavior and be used as an isolator. The detailed consideration of the presented results allows one to draw the same conclusion.

A system is reciprocal between two current sources $\vec{J}_a$ and $\vec{J}_b$ if [1]

$$\iiint_V dV \left( \vec{J}_a \cdot \vec{E}_b \right) \equiv \iiint_V dV \left( \vec{J}_b \cdot \vec{E}_a \right), \qquad (1)$$

where the integral is over all space, and $\vec{E}_a$ and $\vec{E}_b$ are the electric fields that are produced when either the current $\vec{J}_a$ or the current $\vec{J}_b$ is switched on, respectively. Equation (1) states that in order to determine whether or not an optical waveguide is reciprocal with source *a* on one side and *b* on the other one must *project* the electric field generated by one source onto the current of the other source. One cannot simply look at the exiting fields, note they are different, and claim non-reciprocity. If sources *a* and *b* correspond to different modes of the waveguide system, then the difference in the projections would be demonstrated by the inequality of transmission coefficients characterizing forward and backward transmissions between these two modes. If the source *b* additionally excites some mode *c*, this fact, for instance, does not contradict the reciprocity relation between sources *a* and *b*.

The waveguide described in the commented article has two endfaces with two modes each, symmetric (S) and antisymmetric (A). The reciprocity theorem would state that the transmission coefficient from any mode X of endface 1 to any mode Y of endface 2 ($T_{1X->2Y}$) will be the same as its counterpart in the reverse direction, namely, the transmission from mode Y of endface 2 to mode X of endface 1 ($T_{2Y->1X}$). The authors do not compare such transmissions in the main article, instead they stress the difference in transmission between $T_{1S->2A}$ and $T_{2S->1A}$ and refer to it as nonreciprocal light propagation. These transmission coefficients are indeed strongly different. However the different behavior of two such independent transmission paths does not represent non-reciprocity. Thus, an optical isolator cannot be built based on this fact.

Instead the presented data clearly shows reciprocity. For instance, the transmission $T_{1S->2S}$ is reciprocal, as can be seen in figure S1 on page 4 of the supplementary materials section. The red lines in part A and B of the figure demonstrate the transmission coefficients $T_{1S->2S}$ and $T_{2S->1S}$ as functions of waveguide length. Within the computational accuracy and the resolution of the figure the two curves appear to be identical. From the presented data it is not possible to conclude non-reciprocity of transmissions $T_{1S->2A}$ or $T_{2S->1A}$, since their respective counterparts ($T_{2A->1S}$ and $T_{1A->2S}$) are not presented. The transmission $T_{1A->2A}$ and its counterpart are not discussed either. If the authors should have intended to claim non-reciprocity of any of these transmissions, their respective counterpart transmissions should have been presented. From the general perspective of the reciprocity theorem we expect all these transmissions to be reciprocal as well.